# The Tiny Median Filter: A Small Size, Flexible Arbitrary Percentile Finder Scheme Suitable for FPGA Implementation

Jinyuan Wu

*Abstract*—**This document reports the design, implementation and testing of a small silicon resource usage, very flexible arbitrary percentile finding scheme called the Tiny Median Filter. It can be used not only as a median filter in image processing with square filtering windows, but also for applications of any percentile filter or maximum or minimum finder with any size of data set as long as the number of bits of the data is finite. It opens possibilities for image processing tasks with non-square or irregular filter windows. In this scheme, data swapping or data bit manipulating are avoided and high functional efficiency of the logic components is applied to save silicon resources. Some logic functions are absorbed into other functions to further reduce the complexity. The combinational logic paths are designed to be sufficiently short so that the firmware can be compiled to the maximum operating frequency allowed by the block memories of the FPGA devices. The Tiny Median Filter receives, processes and output data in non-stop manner with no irregular timing which helps to simplify design of surrounding stages.**

## I. INTRODUCTION

THE median filter is an essential building block for signal processing in a wide range of fields including high energy physics experiment detector, medical signal processing and image noise reduction. Median filter is a non-liner filter which is significantly more difficult to implement [1-6] than many common liner filters such as sliding average filters. For example, in traditional algorithms, to find the median of 25 data samples, >200 comparisons and data swaps are needed.

Today, the main stream implementations in field-programmable-gate-array (FPGA) of median filter in image processing for 5x5 pixels or other odd number square pixel windows use the scheme developed by Kolte, Smith and Su [1]. The Kolte-Smith-Su scheme requires sorting in columns, partial sorting in rows, partial sorting in diagonals and partial sorting in diagonals with different slope. The total number of comparisons for 5x5 pixel window is reduced to 90, but the implementations still take relatively large amount of silicon resources. Typical implementations require high grade FPGA devices and usually consume > 1000 slices in AMD/Xilinx -7 and higher families. High silicon resource usage brings higher system cost and also causes implementing the median filter in application-specific-integrate-chip (ASIC) very difficult.

For many applications, the data are integers with limited number of bits. For example, in high energy physics detectors, waveforms are usually digitized using 8 to 12-bit analog-to-digital convertors (ADC) while in image processing field, the color or brightness of a pixel is typically an 8-bit integer. Taking advantage of finite number of bits in the data helps to simplify the algorithm such as in the bit-level median filter [6,7].

Going further beyond the bit-level approach, we developed the Tiny Median Filter, a very small size arbitrary percentile filter suitable for FPGA implementation as shown in Fig. 1.

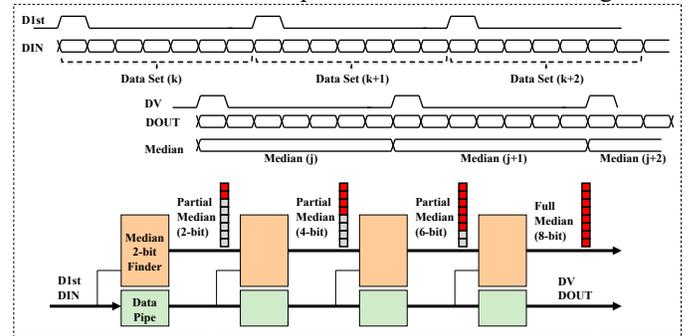

Fig. 1. The operation and structure of Tiny Median Filter.

In the Tiny Median Filter, 2 bits of the input data, instead of 1 bit in bit-level approaches, are examined per stage. The data points are fed into the Median 2-bit Finder block one point per clock cycle. The first block checks the highest 2 bits (bits 7 and 6 for applications with 8-bit data) of each input data point and count number of points with bits 7 and 6 = 11, 10 and 01. Once the entire data set is fed in, highest 2 bits (bits 7 and 6) of the median is determined and output as a partial median. The second stage receives the partial median and determines bits 5 and 4 of the median and subsequent stages determine the lower bits. After 4 stages, the full precision median is determined and output.

These processing stages operate in pipeline manner: as the later stage processes a data set, the earlier stage can process the another data set simultaneously. The first point of a new data sets are fed into the system in the clock cycle just after the last clock cycle of the previous data set so that the entire system matches the speed of the data providing stage.

The features of the Tiny Median Filter includes:

- very small size; The basic version of the Tiny Median Filter uses <200 arithmetic logic modules (ALMs) in an Intel/Altera Cyclone 5 FPGA [8], which is equivalent to <100 slices in the AMD/Xilinx -7 family. Some high throughput versions use <2000 ALMs which is equivalent to <1000 slices.
- accepting arbitrary data sets; The Tiny Median Filter can be used for data sets with any number of points. So the filter can be used for all applications, not limited to image processing. But it also allows the image processing users to explore a much broader range of possibilities such as non-square filter windows with 3x7 or 5x9 pixels, or even with non-rectangular windows.
- non-stopping operation; The input data can be fed into the filter every clock cycle without any stopping. The different data set can be fed end-by-end without wasting a single clock cycle. There is no irregular timing in the operation so it requires no controls with complicate state machines or on-chip microprocessors.
- allowing used as a percentile filter; The users may set a constant M in the Tiny Median Filter to specify the rank of the largest number to be identified in the N data points. So it can be used as an arbitrary percentile filter and the median filter is simply a special case (50%) of the percentile filter.
- allowing used as a maximum or minimum finder;
- suitable for various data precision; For example, if the data from 10-bit ADC are to be processed, simply expand the Tiny Median Filter to 5 stages.

In the remaining parts of this documents, the operating principle is discussed in Section II. The details of implementation of the basic version of the Tiny Median Filter in an Intel/Altera FPGA is described In Section III, followed by testing results presented in Section IV and discussion of the single core performance in Section V. A few high throughput versions are described in Section VI and VII and a special version called the 9753 Tiny Median Filter is presented in Section VIII. The silicon resource usages are presented in Section IX and the document is closed with conclusion in Section V.

## II. OPERATING PRINCIPLE OF THE TINY MEDIAN FILTER

The operating principle of the Tiny Median Filter is based on the quaternary search method. An example of the operation is shown in Fig. 2.

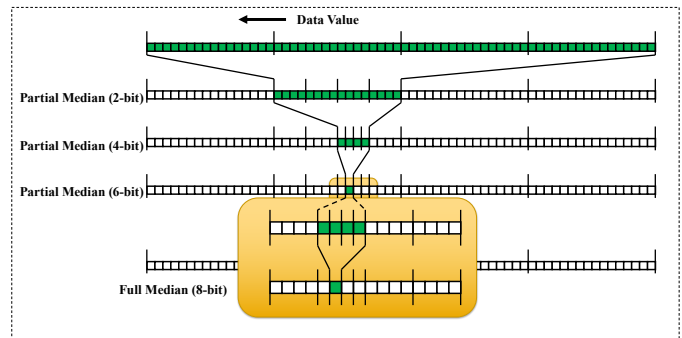

**Fig. 2.** An example of the Tiny Median Filter operation.

Assume we have N points in each data set and each data point is an integer with B bits (we consider 8 bits in the rest of this section). Our goal is to find the M-th largest value within the data set. In each stage of the Tiny Median Filter, a partial median is input which represents a range in which the median exists. The range is subdivided into four and the following two sets of operations will be performed in each stage to further narrow down the range by factor of four:

1. All N data points in a data set are checked, one clock cycle per data point. There are three counters Q3x, Q2x and Q1x in each stage. If the value of the data point is bigger than or equal to the 3 boundaries of the 4 subranges, corresponding counters are increased by 1.
2. After all N data points in the data set are checked, the results of counters Q3x, Q2x and Q1x are compared with M. If Q3x, Q2x and/or Q1x >= M, it means that the median is in one of the ranges, the lowest two bits in the partial median will be set to reflect the comparisons.

In our example shown in Fig. 2, before the first stage, obviously the median can have any value in the entire range, 0 to 255 which is subdivided into 4 subranges. The data points with values >= 192, 128 and 64 are counted, respectively. After checking through all N values, assume the condition Q3x >= M is not true while Q2x and Q1x >= M are true, one can determine that the median is in the range of 128 to 191. The highest two bits of the median, i.e., bits 7 and 6 are then determined = 10 as the result of the comparisons.

In the second stage in Fig. 2, the boundaries are set to be 144, 160 and 176 (128+16, +32 and +48) and number of data points with values >= than these boundaries are accumulated in its counters. Once all N data points are checked, the final counts Q3x, Q2x and Q1x will be compared with M to narrow down the range by a factor of 4. This determines the bits 5 and 4 of the median and the second state will provide to the next stage a partial median containing highest four bits.

In the subsequent stages, the ranges containing the median are further narrowed down by a factor of 4 per stage and the precision of the partial median increases by 2 bit per stage. After the fourth stage, full median with 8 bits precision is determined and output.

It should be pointed out that the computing complexity of this scheme is O(N*log(R)) where R is the range of the data.

Some optimization approaches commonly used in software development are not taken. For example it is a typical practice to eliminate out of orange data in software development that will help to reduce average computing operations. But in our case, it would create various lengths of the data sets, which would need fairly complicate state machines to handle the irregular timing. The irregular timing will create unnecessary challenges to the data input and median receiving stages. Also the variable lengths of the data set would require FIFO memories, rather than fix length circular buffer to hold data points in the data pipeline and that would also increase silicon resources consumption.

Although the algorithm is not optimized in the conventional fashion, the Tiny Median Filter can still achieved small silicon resource usage essentially by using logic functions efficiently. Most components in the circuit perform necessary function such as comparison or addition every clock cycle without pulsing or waiting for data. Some functions are simplified in the design stage. Therefore, although the number of computations is not small, it can still be done with relatively small amount of silicon resources while the data is fetching in.

There is no data swapping or data bit manipulations such as set some data bits to 1 or 0 in our scheme. This will not only save silicon resources, but also provide unspoiled data for the later stage to use.

## III. IMPLEMENTATION DETAILS

The Tiny Median Filter we developed has been implemented in an Intel/Altera Cyclone 5 FPGA. The firmware can be ported to other FPGA families since it only uses general purpose silicon resources such as lookup tables, D-flip-flops, block or distributed memories that can be found in most FPGA devices.

### A. Global Structure

The global structure of the Tiny Median Filter is shown in Fig. 3.

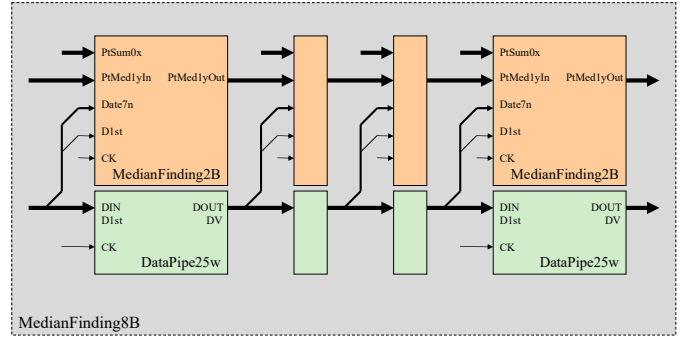

**Fig. 3.** The global structure the Tiny Median Filter.

It contains a chain of 4 median range 2-bit finder blocks marked as "MedianFinding2B" and a chain of 4 data pipe blocks marked as "DataPipe25w".

The external data feeder will send 8-bit data "DIN" along with a first data marker "D1st" to the data pipe (see Fig. 1). The first data marker along with a few bits of the data will be sent to the median finding blocks. The median finding block will check the data and determine two bits of the median.

For example, the first block will use bits 7 and 6 of the data to determine the bits 7 and 6 of the median. The second stage will use bits 7 to 4 of the data to determine the bits 5 and 4 of the median. The third stage will use bits 7 through 2 of the data to determine the bits 3 and 2 of the median and so on.

The partial median will be sent into each median finding block and then sent out with two additional effective bits. At the end of the chain of the 4 blocks, a full precision median with all 8 bits determined will be output.

The partial sum input "PtSum0x" is a constant that can be dynamically changed by a status register or hardware switches. This constant is used by the median finding block to set a comparison condition which will be discussed later.

### B. The Structure of the Median Finding Block

The structure of the 2-bit median finding block can be shown in Fig. 4.

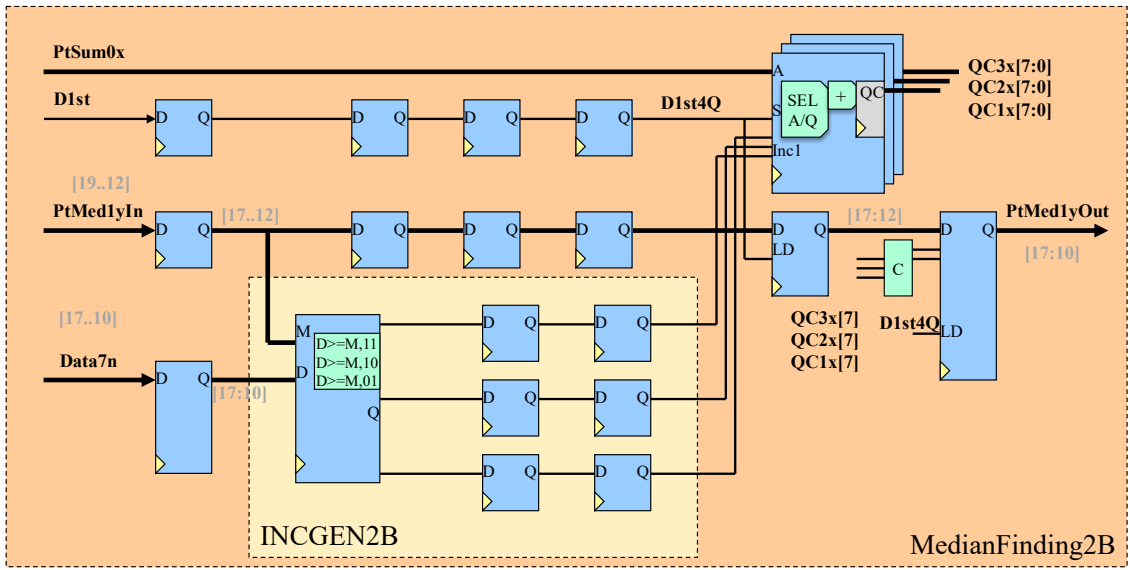

**Fig. 4.** Block diagram of the median finding block.

At the input port of the median finding block, the first data marker D1st becomes HI for one clock cycle. It indicates that a new data set begins and the partial median "PtMed1yIn" from the previous stage is valid. The partial median will remain constant for the entire data set.

The incoming data bits in "Data7n" will be compared with the partial median. If the value of the data point is >= one or several of the 3 boundaries in the range given by the partial median, 1, 2 or 3 incremental signal(s) will be sent out from the block INCGEN.

Each of the incremental signals will be sent to one of the 3 counters with a input port marked "Inc1", and when Inc1=HI, the counter value will increase by 1.

Once entire N data points are fetched through, the counter values are checked with the conditions Q3x >= M, Q2x >= M and Q1x >= M. These conditions are represented by QC3x[7], QC2x[7] and QC1x[7] as shown in the block diagram. These comparison results feed through a lookup table marked with "C" to determine two additional effective bits of the partial median which will be output via the " PtMed1yOut" port.

### C. Incoming Data Comparing Operation

An example of the incoming data comparing is shown in Fig. 5.

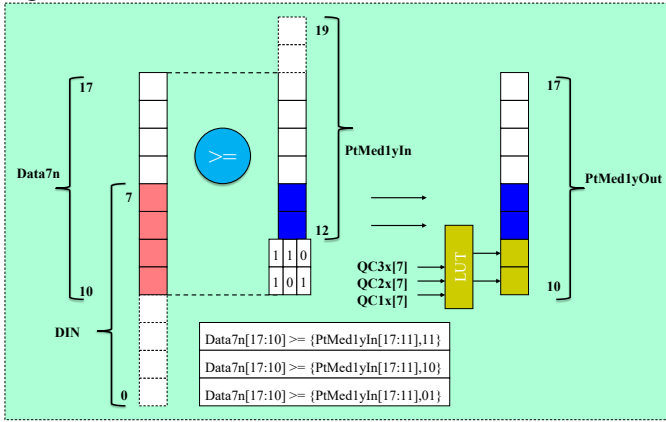

**Fig. 5.** An example of the incoming data comparing.

The bits from the data are brought the median finding block to port "Data7n". The first stage will only need the highest two bits, i.e., bits 7 and 6 of the data, while the subsequent stages need two additional bits. For example, as shown in Fig. 5, the second stage will need bits 7 to 4 of the incoming data "DIN" and they are wired to Data7n[13:10]. The partial median input "PtMed1yIn" of the second stage has two effective bits from the earlier stage. The two effective bits are actually bits 7 and 6 of the final median but they are routed as PtMed1yIn[13:12]. The data bits are compared with the partial median bits concatenated with two bits representing 3 subrange boundaries. The actual logic functions can be described as:

```
Data7n[13:10] >= { PtMed1yIn[13:12], 11}
Data7n[13:10] >= { PtMed1yIn[13:12], 10}
Data7n[13:10] >= { PtMed1yIn[13:12], 01}
```
where the 2-bit partial median is concatenated with 11, 10 and 01 as the lowest bits, respectively.

As shown in the block diagram in Fig. 4, we reserved four pipeline steps to accommodate possible future extension needing complicate logics such as in high throughput versions. The practical implementation demonstrated that the comparison logic alone shown above can be implemented with just one step of the pipeline without degrading operating speed significantly.

Note that the two new effective bits are inserted into the partial median as the lowest bits and as the partial median passes a stage, all bits are shifted up by 2. At the end of the chain of 4 median finding blocks, the lower bits are all shifted up and the output of PtMed1yOut[17:10] will contain 8 bits of full precision median.

Although the comparators used in different stage process different numbers of bits, they are uniformly coded as 8-bit comparators in design stage for convenience. The higher bits of the comparator inputs Data7n and PtMed1yIn are filled with 0's (represented as higher white boxes in Fig. 5) and the compiler will automatically synthesis these excessive bits away.

### D. Automatic Comparison with Counter Initiation

Three counters are implemented that will count numbers of data points whose value is higher than the corresponding boundaries. Once all the N points in the entire data set are checked, the results of the counters Q3x, Q2x and Q1x are to be compared with M: Q3x >= M, Q2x >= M and Q1x >= M. This would need a set of comparators that would take some silicon resource but will not be used efficiently since they are only used once at the end of the counting.

These comparison functions can be absorbed into the counters (with outputs marked QC3x[7:0], QC2x[7:0] and QC1x[7:0]) in our design. At the first clock cycle of a data set, each counter is initiated with a constant PtSum0x = 128 - M. The values of the counters after the first clock cycle are (128 - M + Inc1). (Note that should we implement these counters in regular manner, they would be preset with 0 + Inc1 at the first cycle). Therefore, when we have M or more points counted in the N clock cycles while checking the entire data set, the MSBs of the counters, QC3x[7], QC2x[7] and QC1x[7] will become 1. This way, the comparison is automatically performed without extra silicon resources. The values of QC3x[7], QC2x[7] and QC1x[7] can be used to generate the two new effective bits of the partial median as shown in Table I.

Table I
Truth Table of the Partial Median

| QC3x[7] | QC2x[7] | QC1x[7] | PtMed1yOut[11:10] |
|---------|---------|---------|-------------------|
| 0 | 0 | 0 | 00 |
| 0 | 0 | 1 | 01 |
| 0 | 1 | x | 10 |
| 1 | x | x | 11 |

In the truth table, "x" means that the corresponding input can be either 0 or 1, but in our case, these values should be 1. For example, if Q3x[7] = 1, it means that the median value is >= than the boundary between subregions 3 and 2, which implies that the median value is >= than the boundaries between subregions 2 and 1 and between 1 and 0, yielding Q2x[7] = 1 and Q1x[7] = 1.

It should also be pointed out that only 3 counters are needed to determine the place of the median in the 4 subregions, although at the early design planning stage, it is natural to consider assigning 4 counters, one for each subregion. Implementing 3 rather than 4 counters reduces non-negligible amount of silicon resources.

### E. Data pipe implementation

There are two options to implement the data pipe: using block RAM or using distributed memories. The memories are configures as the simple dual-port mode with a write and a read port. The write address and the read address are provided by two continuous incrementing counters with a constant offset. All the four data pipe blocks have identical depths and therefore they can share the same set of write address and read address counters. In the FPGA devices (Intel/Altera Cyclone 5) that we did the test, the block RAM, M10K can be configures as 40 bits wide by 256 words. To implement 4 blocks of data pipe, one will need just one M10K blocks. With M10K blocks implemented as dada pipe, the data set size N can be as large as about 250. Therefore, for Tiny Median Filter, the data set size N (or filter window size) within certain range makes no difference in silicon resource consumption, which is very different from many main stream median filter design.

The distributed memories are MLABs in Cyclone 5 FPGA, which are actually the lookup tables (LUTs) of the arithmetic logic modules (ALMs). Each LUT can be used as a 32-bit memory which is very convenient for applications with small data set size (for example, for 5x5 pixels filter, N=25).

## IV. TEST RESULTS

The Tiny Median Filter has been tested in an evaluation module (Terasic C5G) [9] with 275 MHz system clock. The

275 MHz clock frequency is restricted by the memory blocks of the FPGA device. The logic paths are designed sufficiently short so there are no setup or hold timing issues are reported by the compiler. The test is in the memory-to-memory fashion as shown in Fig. 6.

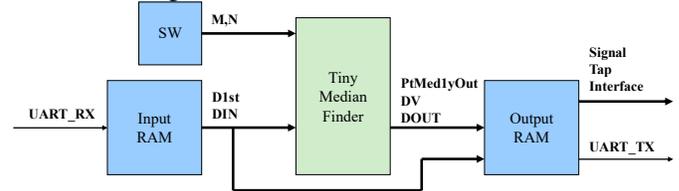

**Fig. 6.** The testing environment.

The input RAM contains data either preloaded or reloaded via the universal asynchronous receiver/transmitter (UART) interface. The evaluation module converts the UART signals to USB interface that can be connected to the test control computer.

The constants N and M are set with the on-board switch which specify the data set size and the rank of the value to be found. For example, when M=13 and N=25, the 13-th highest value in 25 data points are to be found, which is the median of the data set.

The Tiny Median Filter processes input data at 275 MHz clock frequency. The median results PtMed1yOut and the delayed version of the first data marker and the delayed raw data DV and DOUT are output from the Tiny Median Filter and are sent to the output RAM. The non-delayed original first data marker D1st and raw data DIN are also stored in the output RAM for analysis purpose.

The contents in the output RAM can be sent out via URAT interface or via the Signal Tap Interface using USB cables to the test control computer.

### A. Median Finding Operation

The data stored in the output RAM are plotted using the Signal Tap Interface as shown in Fig. 7.

In the top pane, the raw data of 8 data sets, 3 data points each are sent out of the input RAM in a burst. After some clock cycles, the median are shown along with the delayed raw data DOUT and data valid marker DV. A median is output as soon as all N data points of an entire data set are fed through the last stage of the median finder.

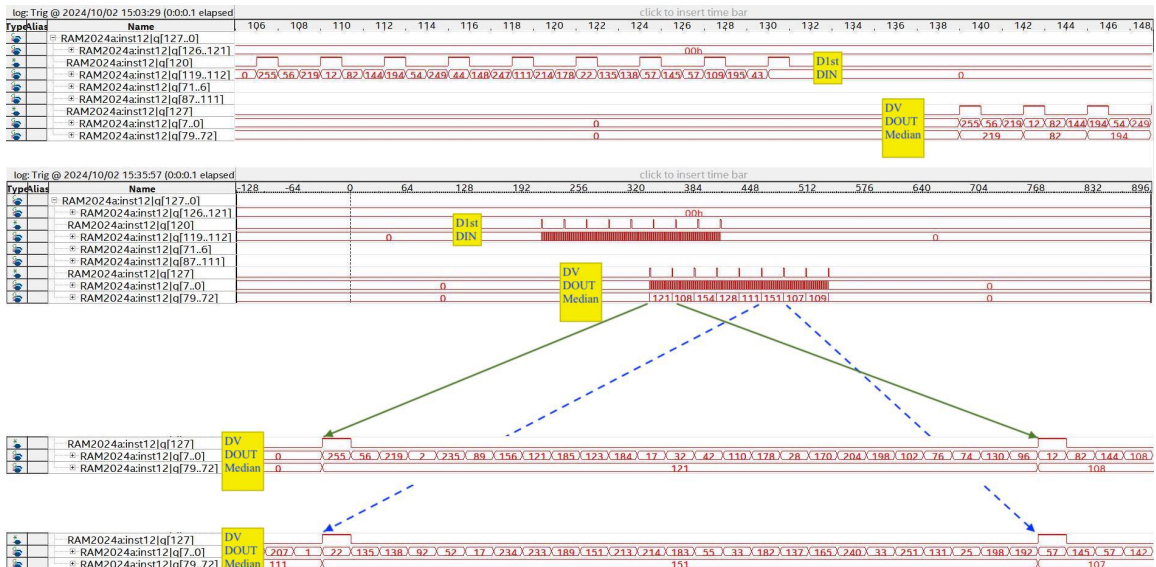

**Fig. 7.** The operation of finding medians for data sets with various data points.

It can be seen that there is no gap between two data sets. The processing rate matches the data feeding rate so that the data feeding is non-stopping.

In the second pane, 25 data points per set are processed with the same Tiny Median Filter. The median value outputs are aligned with DV and DOUT as shown. Two detailed views are shown in the third and the fourth panes.

### B. Flexibilities of Applications

As we pointed out earlier, the Tiny Median Filter actually finds an M-th highest value in N data points. Both N and M are adjustable with a status register or hardware switches without recompiling the firmware. The flexibility of the applications is demonstrated in Fig. 8.

The data set size N is set = 9 and the rank M is set: M = 5 for median, M = 1 for maximum, M = 2 and M = 3 for the 2nd and the 3rd highest value in the data set and M = 9 for minimum. The same data sets are sent through the Tiny Median Finder for each case. Correct outputs are seen in the test.

## V. SINGLE CORE PERFORMANCE OF THE TINY MEDIAN FILTER

The Tiny Median Filter can be used to find any M-th highest value in any data sets with N data points. This flexibility allows users to dynamically adjust filter window size and percentile of the filter output. In the situation of using M10K to implement the data pipe and counters being 8 bits, the medians with up to 250 data points can be found (N=250, M=125) without recompiling the firmware. (In our implementation, the maximum length of the data pipe is 255. The pipeline in the median finding circuits has 5 steps. So the

maximum data set size is 250.)

In image processing applications, the Tiny Median Filter opens unique possibilities of using irregular filter windows. The filter windows need not to be restricted to squares and rectangular or diamond shaped windows are allowed, which provides various options to the users. Several examples of irregular filter windows and the single core performance are shown in Table II.

Table II
Application Examples and Single Core Performance of the Tiny Median Filter

| Operating Frequency | | 275 MHz |
|---|---|---|
| Image Size | | 1024 x 768 |
| Filter Window Shape | Data Set Size N | Single Core Frame Rate |
|  3x3 | 9 | 38.8 (fps) |
| 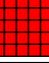 5x5 | 25 | 13.9 (fps) |
|  3x5 | 15 | 23.3 (fps) |
|  3x7 | 21 | 16.6 (fps) |
| 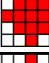 Diamond 5 | 13 | 26.8 (fps) |
| 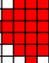 Diamond 7 | 25 | 13.9 (fps) |

To send data to the Tiny Median Filter for irregular filter window applications such as Diamond 5 or 7 windows, simply generate addresses of the memory that store the pixel data. For example, for Diamond 7 window, one needs to implement three counters, one for X and Y coordinates each and the third 5-bit counter for 25 pixels. The third 5-bit counter will feed into two lookup tables that will produce X and Y offsets for each of the 25 pixels. The offsets are added with the X and Y counter outputs to produce the address for the pixel data memory.

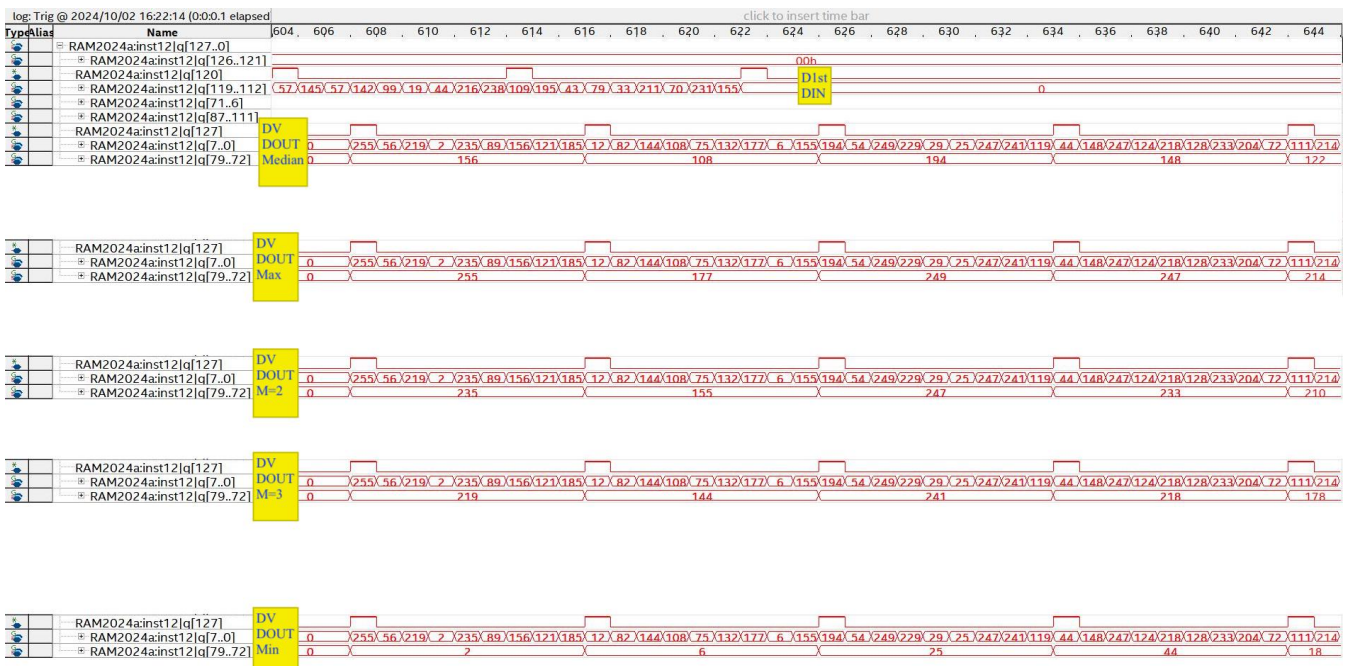

**Fig. 8.** The operation of finding medians, maximum, the 2nd, the 3dr highest and the minimum for data sets with 9 data points.

The frame rates shown above already meet requirements of many applications. When larger images with higher frame rate are to be processes, high throughput versions of the Tiny Median Filter will be needed, which will be discussed in the next sections.

## VI. MULTI-CHANNEL VERSION OF THE TINY MEDIAN FILTER

To increase frame rate or/and process larger images, a natural approach is to duplicate hardware so that multiple cores can process the data parallelly. Directly duplicate the single core certainly will meet the throughput requirement but it is less efficient in terms resource usage.

In our design, the processing functions are partitioned carefully so that if multiple cores are to be implemented, some logic functions can be shared instead of duplicated. So the silicon resource usage in the high throughput versions of the Tiny Median Filter is not directly proportional to the single core one, but much smaller.

The throughput of the Tiny Median Filter can be improved as long as the data input stage can provide data with sufficiently high rate. Inside FPGA, there exist large number of small block memories that can be used as feeders to the Tiny Median Filter. Consider a possible structure of "pixel strip buffer" as shown in Fig. 9 which stores pixel data of a strip in an large image.

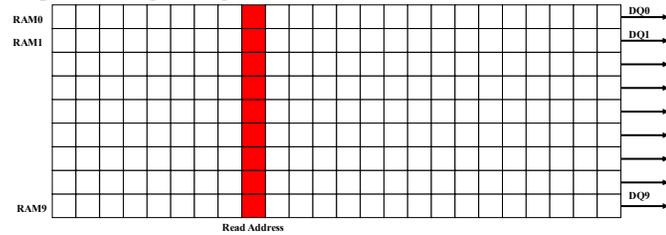

**Fig. 9.** The "pixel strip buffer"

Each row in the strip is implemented with a block RAM that provide a pixel data point every clock cycle. Data from multiple rows are sent to the multi-channel Tiny Median Filter. The block diagram of a 9-channel version of the median finding block is shown in Fig. 10.

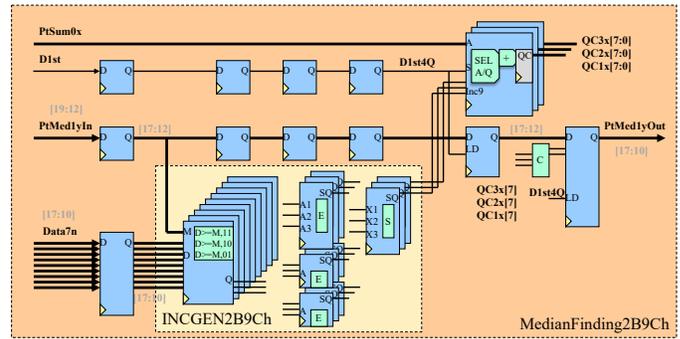

**Fig. 10.** The multi-channel median finding block.

The basic structure of the multi-channel median finding block is similar as the single channel one. In the "INCGEN2B9Ch" block, input data from 9 channels are compared with 3 boundaries, respectively, that produce 27 result bits of comparisons. These 27 bits are first feed to a set of 3-in-2-out encoders marked with "E" and the results are 9 2-bit integers representing how many of the comparison results are "true". These 9 2-bit integers are summed by 3 adders marked with "S" in the next pipeline stage. Each of the adders accept 3 2-bit integers and add them together to produce a 4-bit sum.

The counters in the single-channel version now become accumulators with input "Inc9" being a 4-bit integer. The remaining logic functions are identical as in the single-channel version.

It can be seen that large portion of the logic functions in the multi-channel version have just one copy. It would need multiple copies of them should we simply duplicate single-channel version for high throughput applications.

The multi-channel version of the Tiny Median Filter processes one column of data per clock cycle. For example, if we need to find median of pixels in 9x9 windows, it will need 9 clock cycles with a 9-channel Tiny Median Filter.

The multi-channel version still support some (but not all) flexibilities as the single-channel version of the Tiny Median Filter. For example, the multi-channel version can still be used as a percentile filter and the filter windows can also be a rectangle. An example is shown in Fig. 11.

In the Signal Tap display, "q[120]" is the first data marker "D1st", followed with a few lines representing the input data "DIN". There are 9 input channels but we only displayed 5 due to limit of the output memory data port width. The bit "q[127]" is the output data valid signal "DV" and the 9 lines below represent the 9 channels of output data "DOUT".

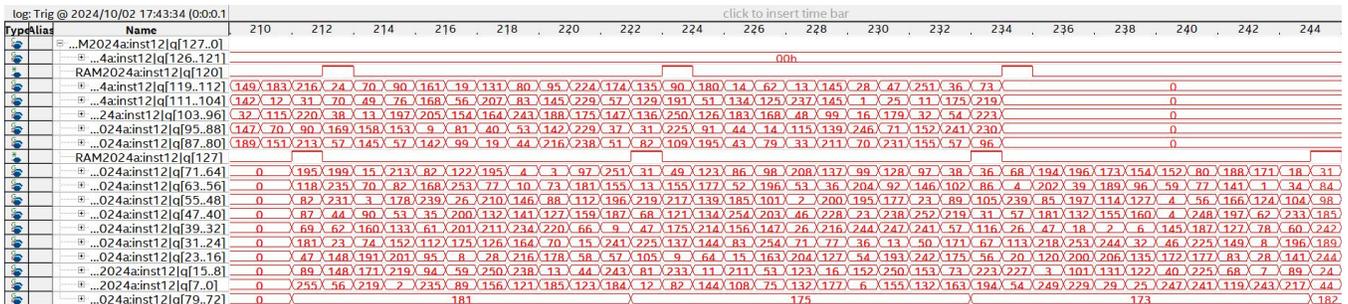

**Fig. 11.** The operation of finding the 31st highest values in 9x11 windows.

The last line in the diagram shows output results from the Tiny Median Filter.

In this example, the filter window size is 9x11 and M = 31, which means to find the 31st highest in the 99 values within the window and correct results are seen. (To find the median in 99 values, simply set M = 50).

## VII. SINGLE-CYCLE VERSION OF THE TINY MEDIAN FILTER

In some applications, very high frame rates are needed so that the median in an N x N window must be generated every clock cycle. To fulfill this requirement, multiple multi-channel Tiny Median Filter described in previous section are to be duplicated.

In our design, input pixel data are stored in the strip buffer as discussed in previous section but they are examined by multiple filters in the sliding window fashion and an example with 9x9 windows is shown in Fig. 12.

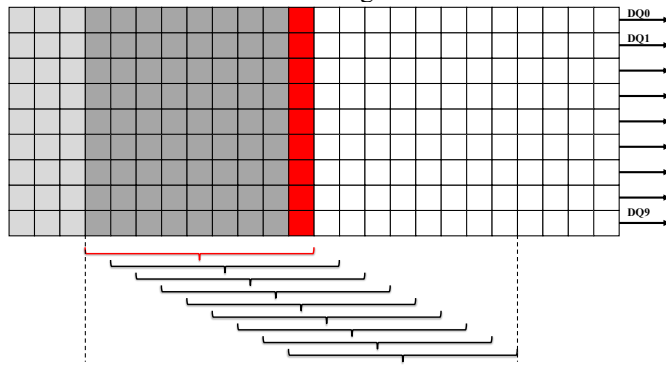

**Fig. 12.** The sliding filter windows.

The data are readout from the strip buffer one column per clock cycle and the column address steps from left to right. At a given clock cycle, the data feed into 9 filters. These 9 filters have different starting points of the data window and at each clock cycle, the median in one of these 9 filters becomes mature. This way, the entire filter ensemble has one median ready every clock cycle and the mature median can be routed to the output for the next stage.

The global structure of the single-cycle version of the Tiny Median Filter is shown in Fig. 13.

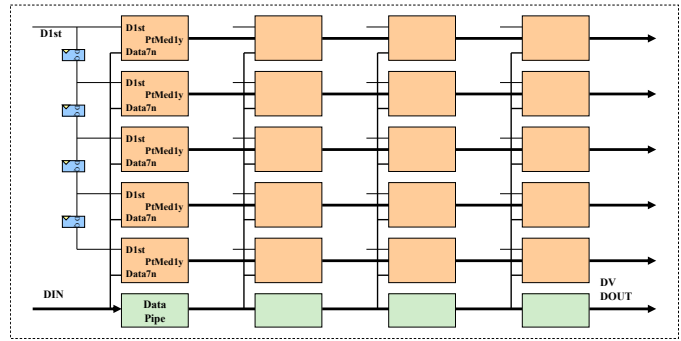

**Fig. 13.** The single-cycle version of the Tiny Median Filter.

The single-cycle version of the Tiny Median Filter consists of several chains of the multi-channel median finding blocks and a chain of the data pipe. For example, for the 5x5 filter window, 5 median finding block chains are needed with each block being the 5-channle one and for 9x9 window version, 9 chains of 9-channel blocks are needed.

The blocks in each chain examine identical data and this is the reason why only one data pipe is needed. The primary difference between each median finding block chain is that their first data markers "D1st" have different timing. As shown in the block diagram, the first data marker "D1st" feeds the first chain directly and is delayed by one clock cycle to feed the second chain. The subsequent blocks are fed with delayed version and this arrangement causes different chains to interpret the data with sliding windows.

The arrangement like this not only saves silicon resource for the data pipe, but also saves some logic functions in the median finding blocks. The comparison logics in the first blocks of different chains are identical and the compiler will automatically synthesis repeating logics away.

The single-cycle version of the Tiny Median Filter outputs one result per clock cycle. Typical variations include 9x9, 7x7, 5x5 and 3x3 filter windows and their silicon resource usages are different.

The single-cycle version still support some (but not all) flexibilities as the single-channel version of the Tiny Median Filter. For example, the single-channel version can still be used as a percentile filter and an example is shown in Fig. 14.

In the Signal Tap display, bit "q[120]" is the first data marker "D1st", followed with a line of data representing the input data "DIN". There are 9 input channels but we only displayed one due to limit of the output memory data port width. The bit "q[127]" is the output data valid signal "DV" and the 9 lines below represent the 9 channels of output data

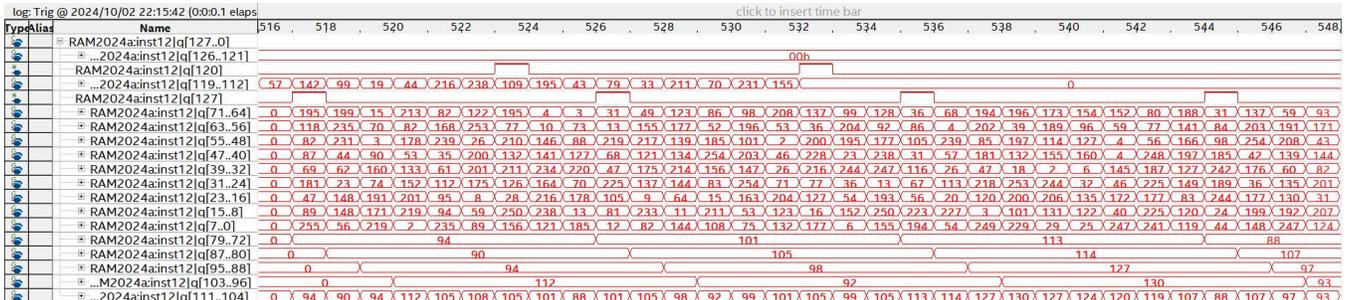

**Fig. 14.** The single cycle output of the 48th highest value in the 9x9 sliding windows.

"DOUT". The lines marked with "q[79..72]", "q[87..80]", "q[95..88]" and "q[103..96]" are first 4 output results (There are 9 but only 4 are shown). The last line in the diagram shows the combined output results from all the 9 chains with one result per clock cycle.

In this example, the filter window size is 9x9 and M = 48, which means to find the 48th highest in the 81 values within the window and correct results are seen. (If the medians are needed, set M=41).

## VIII. THE 9753 TINY MEDIAN FILTER

In some applications, filtering with several different window sizes may be needed. For example, one may need filter results of an image for 9x9, 7x7, 5x5 and 3x3 windows at the same time as shown in Fig. 15.

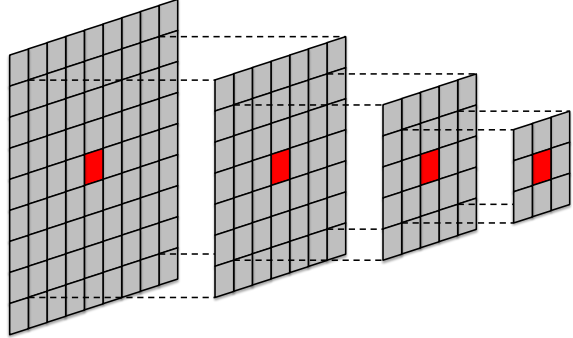

**Fig. 15.** The 9x9, 7x7, 5x5 and 3x3 filter windows for same set of pixels

It is definitely possible to put the 9x9, 7x7, 5x5 and 3x3 filters together to fulfill this requirement but we can also design them together for a optimal balance for small silicon resource usage while meet the operation requirements.

We call this variation the 9753 Tiny Median Filter and its block diagram is shown in Fig. 16.

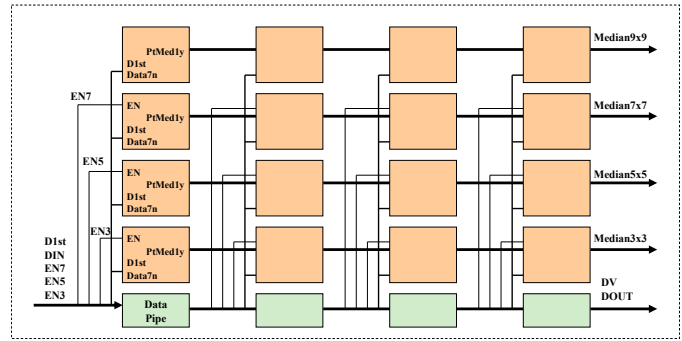

**Fig. 16.** The 9753 Tiny Median Filter

The 9753 Tiny Median Filter consists of several chains of the multi-channel median finding blocks and a chain of the data pipe. The first chain is constructed with the 9-channel median finding blocks and the remaining chains are constructed with 7-channel, 5-channel and 3 channel blocks.

In the 9753 Tiny Median Filter, one column of data are processed per clock cycle. Therefore, to find median of pixels in 9x9 windows, it will need 9 clock cycles with the 9-channel chain. Within 9 clock cycles, other medians for 7x7, 5x5 and 3x3 windows will also be calculated. Enable signals EN7, EN5 and EN3 are generated by the external control circuit and are applied to the 7-channel, 5-channel and 3-channel chains, respectively. The EN7 signal will let the 7-channel chain to ignore first and last columns and process only 7 columns in the middle and EN5 and EN3 will have similar functionality.

The 9753 Tiny Median Filter can still be used as a percentile filter, and the different chain can have different percentile setting, although in most applications, users will choose regular medians. An example of the 9753 Tiny Median Filter operation is shown in Fig. 17.

In the Signal Tap display, bit "q[120]" is the first data marker "D1st", followed with a line of data representing the input data "DIN". There are 9 input channels but we only displayed one due to limit of the output memory data port width. The bit "q[127]" is the output data valid signal "DV" and the 9 lines below represent the 9 channels of output data "DOUT". The lines marked with "q[79..72]", "q[87..80]", "q[95..88]" and "q[103..96]" are the medians for the 9x9, 7x7, 5x5 and 3x3 windows. The bit "q[106]", "q[105]" and "q[104]" are the delayed version of the enable signals EN7, EN5 and EN3.

In this example we can see that the 9x9, 7x7, 5x5 and 3x3 medians are produced every 9 clock cycles.

If the users wish, medians with non-square filter windows such as 7x9, 5x9, 3x9 can also be generated by changing the timing of the enable signals EN7, EN5 and EN3.

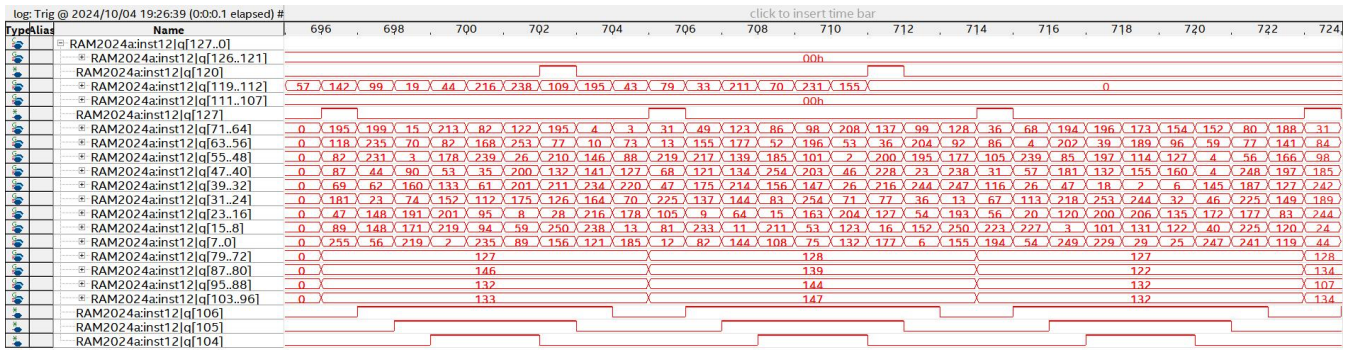

**Fig. 17.** The operation of the 9753 Tiny Median Filter

## IX. SILICON RESOURCE USAGE

In the design scheme of the Tiny Median Filter, logic functions are performed efficiently and unnecessary data manipulations are avoided. Therefore all variations of the Tiny Median Filter use significantly less silicon resources than main stream implementations today.

Several versions of the Tiny Median Filter described in previous sections have been implemented in an Altera Cyclone 5 FPGA (5CGXFC5C6F27C7N). For the single-core Tiny Median Filter, two versions, one using M10K block RAM and the other using MLAB distributed memories, respectively, for the data pipe, are compiled. The resource usages are shown in Table III.

Table III
Resource Usages of the Tiny Median Filter

| Version | Intel/Altera Cyclone 5 FPGA | | AMD/Xilinx -7 family FPGA (estimate) | | CLK Cycles Per Median | CLK FRQ. (MHz) |
|---|---|---|---|---|---|---|
| | ALM | M10K | Slices | RAMB | | |
| 1-Core (M10K) | 83 (130) | 1 | <70 | | 25 for 5x5 window | 275 |
| 1-Core (MLAB) | 159 (235) | 0 | <82 | 0 | 25 for 5x5 | 275 |
| 1x9 | 253 (358) | 8 | | | 9 for 9x9 | 275 |
| 3x3 | 325 (436) | 3 | | | 1 for 3x3 | 275 |
| 5x5 | 823 (990) | 5 | | | 1 for 5x5 | 275 |
| 7x7 | 1307 (1568) | 7 | | | 1 for 7x7 | 275 |
| 9x9 | 1959 (2295) | 9 | <1000 | 5 | 1 for 9x9 | 250 |
| 1x9753 | 699 (867) | 10 | <400 | 5 | 9 for 9x9, 7x7, 5x5, 3x3 | 250 |

The ALM and the M10K usages for Intel/Altera Cyclone 5 FPGA are taken from actual complier reports. The numbers in parentheses used are the numbers of ALMs placed, in which some ALMs are only partially used and the unused resource usually can be reclaimed for other logic.

The slice and block memory (RAMB36E1) usages for AMD/Xilinx -7 family devices are hand guided by the Cyclone 5 compiling results. The resource usages for AMD/Xilinx -7 family devices are calculated here as reference for readers who are familiar with AMD/Xilinx FPGA families only. Usually a slice has more than x2 of logic resources than a ALM and a slice usually can implement logic functions that need two ALMs. But we choose more conservative estimates without considering the optimization by the compiler.

When the block memory is replaced with distributed memory, 72 additional ALMs are used for the data pipe,

assuming the maximum length of data pipe is less than 32. The compiler may have different efficiency in other logic blocks so the difference shown above is not exactly 72.

In most versions of the Tiny Median Filter, we have achieved an operating frequency of 275 MHz, which is the highest frequency allowed by the M10K block memories in this device. The 9x9 and the 1x9753 versions are compiled with a slightly slower operating frequency of 250 MHz to reduce difficulties for the compiler to automatically find acceptable placements.

## X. CONCLUSION

The principle, implementation and testing results of the Tiny Median Filter are discussed in this document. The Tiny Median Filter can be used to find any M-th highest value in any data sets with N data points. This flexibility allows users to dynamically adjust filter window size and percentile of the filter output. The small silicon resource usage enables users to implement more applications in FPGA or ASIC, which traditionally can only be implemented with software.